\documentclass[a4paper,11pt]{article}
\usepackage{pos}
\usepackage{dcolumn}
\usepackage{bm}
\usepackage{amsmath}
\usepackage{graphicx,color} 
\usepackage{subcaption}

\begin{document}
\title{Flux tube formation and the Weingarten representation of center vortices and chains}
\author[a]{David R. Junior}
\author*[b]{Luis E. Oxman}
\affiliation[a]{Instituto de F\'isica Te\'orica, Universidade Estadual Paulista and South-American Institute of Fundamental Research ICTP-SAIFR, Rua Dr. Bento Teobaldo Ferraz, 271 - Bloco II, 01140-070 São Paulo, SP, Brazil}
\affiliation[b]{
Instituto de F\'\i sica, Universidade Federal Fluminense, 24210-346 Niter\'oi, RJ, Brasil.}
\emailAdd{david.rosa-junior@unesp.br}
\emailAdd{leoxman@id.uff.br}
\abstract{We review some recent results regarding the formulation of mixed ensembles of oriented and nonoriented center vortices  based on the Weingarten representation for the sum over surfaces. This framework enabled a partition function for the Abelian-projected ensemble of vortex worldsurfaces, previously formulated as a wavefunctional peaked at center-vortex loops. In particular, we showed how an Abelian ensemble can account for $N$-ality while supporting a ``dual superconductor'' model for confinement. This formulation also clarified the description of the Goldstone modes for  percolating surfaces with non-Abelian degrees of freedom, used in the original mechanism for the formation of a confining flux tube due to the percolating mixed ensemble.} 

\FullConference{The XVIth Quark Confinement and the Hadron Spectrum Conference\\
Cairns, Queensland, Australia \\
18–24 Aug 2024 
}
\maketitle 
\section{Introduction}

 Among the most prominent approaches to confinement in $SU(N)$ Yang-Mills theory, center vortices have emerged as fundamental configurations that capture essential aspects of the infrared dynamics \cite{Mandelstam1976}-\cite{MF}. These structures naturally account for the $N$-ality properties observed in lattice simulations \cite{n-ality}-\cite{cv-11}. However, center-projected ensembles do not explain the formation of a flux tube. Another important line of research was initiated in Ref. \cite{superconduct-1}, based on Abelian-projected fields and the associated monopoles, which were  detected by means of the DeGrand-Touissant method \cite{touissant}. While Abelian-projected scenarios successfully capture $N$-ality and reproduce the flux tubes between quarks at asymptotic distances,  monopole-only approaches are not compatible with $N$-ality \cite{d-wl}. To address these limitations, it was proposed that mixed ensembles of percolating center vortices and chains  can integrate the $N$-ality properties and the  formation of flux-tubes, offering a unified description of confinement \cite{mixed}. Chains are collimated configurations, nonoriented in the Lie algebra, formed by elementary center vortices with distinct magnetic weights interpolated by a monopole. Then, a lattice model of non-Abelian link-variables $U_\mu \in SU(N)$ with frustration was developed to represent the average of center elements. The model naturally incorporates the matching of $N$ vortex worldsurfaces ($N$-matching). In particular, based on Ref. \cite{Rey},  it was argued that these variables can be thought of as the Goldstone modes in a center-vortex condensate with non-Abelian degrees of freedom (d.o.f.). Nonoriented configurations were introduced via an ensemble of adjoint holonomies. Using polymer techniques, this sector led to effective adjoint Higgs fields. When the monopoles on chains also condense, the correct spontaneous symmetry breaking $SU(N) \to Z(N)$ for the formation of flux tubes with $N$-ality was established. Later, following this line of thought, oriented and nonoriented center vortices were modeled within an Abelian-projected framework, using a wavefunctional localized around the corresponding gauge fields \cite{wavefunctional}. In the electric field representation, polymer methods were employed to integrate over the ensemble and express the wavefunctional in terms of an effective field theory. The ensemble was constructed using elementary center vortices carrying Cartan fluxes $\beta_i$. In the monopole sector, the simplest elementary charges $\beta_i - \beta_j$, corresponding to the magnetic weights of the adjoint representation, were considered. In this manner, important properties of confinement were understood in an Abelian-projected setting.
 
A central point we addressed in Ref. \cite{weingarten-new}, based on the Weingarten matrix representation \cite{weingarten} of the partition function, is how Abelian projection can account for $N$-ality while simultaneously enabling a ``dual superconductor'' model for the fundamental string. Another key aspect we examined within this representation is the emergence of non-Abelian gauge fields as the Goldstone modes for percolating surfaces with non-Abelian d.o.f. In these proceedings, we briefly review these important points regarding the confinement mechanism based on percolating center vortices and chains.

\section{Relevant vortex and monopole configurations}
\label{relevant}

\begin{figure}[h]
    \centering
 \includegraphics[ scale=.23]{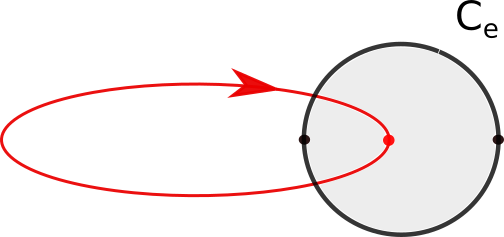}     
 \hspace{1.4cm}      \includegraphics[scale=.23]{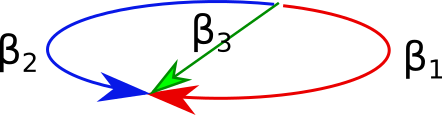}
  \hspace{1.4cm}  
 \includegraphics[scale=.23]{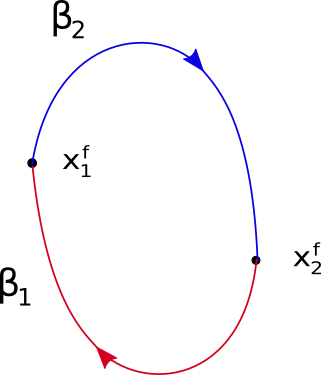}
 \caption{From left to right: a vortex loop linking a Wilson loop; vortex $N$-matching in $SU(3)$; nonoriented chain (with unobservable Dirac strings omitted).}
 \label{vortices-3d}
\end{figure}

This section summarizes key physical aspects of mixed ensembles composed of center vortices and chains in four-dimensional Yang-Mills (YM) theory.  Center vortices form a percolating cluster in the confining phase \cite{percolating-2}. In the continuum, they can be described by singular mappings combined with smooth profiles. For elementary vortices with center charge $± 1$, outside the flux localization region for thick center vortices, the gauge field $A_\mu^{\rm thick}$ takes the form
\begin{align}
    a_\mu = \beta\cdot T \partial_\mu\chi \makebox[.5in]{,} \beta\cdot T \equiv \beta|_q T_q\;,\label{def-cv}
\end{align}
where $\chi$ is multivalued around the vortex guiding-center $\mathcal{S}$, with $\Delta\chi=2\pi $. Moreover, $T_q$ are the Cartan generators of  $\mathfrak{su}(N)$ and $\beta=2N\omega $ is one of the magnetic weights $\beta_i$, $i=1, \dots, N$ of the defining representation of $SU(N)$. This gauge field can also be described in terms of a singular map $S$:
\begin{align}
   {\rm Ad}(a_\mu) =i{\rm Ad}(S)\partial_\mu {\rm Ad}(S)^{-1}\makebox[.5in]{,} S=e^{i\chi \, \beta\cdot T} \;,
    \label{Scv}
\end{align}
where Ad stands for the adjoint representation of the group. When a thick center vortex is fully encircled by a loop $\mathcal{C}_{\rm e}$, the group holonomy can be evaluated using $a_\mu$ instead of $A_\mu^{\rm thick}$ 
\[
{\rm D} \left( e^{i \int_{\mathcal{C}} dx_\mu\, a_\mu(\mathcal{S})} \right) = {\rm D} \left( e^{i \Delta \chi \, \beta \cdot T} \right)=e^{-\frac{2\pi k}{N}L(\mathcal{C}_{\rm e},\mathcal{S})} \;,
\]
where ${\rm D}(\cdot)$ is a general group representation of $SU(N)$,   $L(\mathcal{C}_{\rm e},\mathcal{S})$ is the linking number between $\mathcal{C}_{\rm e}$ and $\mathcal{S}$, and $k=0,\dots, N-1$ stands for the $N$-ality of the representation. 
Due to the property $\beta_1 + \dots + \beta_N = 0$, $N$ vortex lines carrying different magnetic weights can be matched at a common endpoint ($N$-matching). In addition to configurations containing only vortices, there are also chains, which are formed by collimated center-vortex fluxes attached in pairs to monopoles. These configurations were detected in lattice simulations for both $SU(2)$ and $SU(3)$ \cite{collimation,Stack2002}. These correlations are shown on different $\mathbb{R}^3$ slices in Figs. \ref{vortices-3d} and  \ref{vortices-4d}. Monopole worldlines carrying different adjoint weights can also match at a point in $\mathbb{R}^4$. For instance, in $SU(3)$, three monopole worldlines with charges satisfying $\alpha + \gamma + \delta = 0$ can meet at endpoints, forming a closed configuration (see Fig. \ref{fig:main}). Higher-order matchings are also possible.

\section{ Abelian-Projected  partition function}
\label{wavefun}

One objective of Ref. \cite{weingarten-new} was to represent, at the level of the partition function, the Abelian-projected ensemble considered at the level of the vacuum wavefunctional in Ref. \cite{wavefunctional}.  In this section, we briefly review the main steps followed in that reference. Initially, we considered configurations containing no monopoles (oriented in the Lie algebra) and a  functional peaked at gauge fields $a(\{\gamma\})$, where $\{ \gamma\} $ denotes a particular vortex network formed by vortex lines with a distribution of magnetic weights. For elementary center vortices (center charge $\pm 1$), there are $N$ possibilities $\beta_i$.  The gauge field associated to each vortex line is given by
\begin{align}
 a_{\beta}(\gamma) = 2\pi\beta \cdot T \,(-\nabla^2)^{-1}\nabla \times j(\gamma)\makebox[.5in]{,}j(\gamma) = \int_{\gamma} d\bar{x}\, \delta(x - \bar{x}) \;.
\end{align}
When $\gamma$ is a loop, $a_\beta(\gamma)$ represents the gauge field of a thin center vortex with guiding-center $\gamma$ and magnetic charge $\beta$. The associated Wilson loop is given by $e^{-\frac{2\pi k}{N}L(\gamma,\mathcal{C}_{\rm e})}$, where  $L(\gamma,\mathcal{C}_{\rm e})$ is the linking number between $\gamma$ and  $\mathcal{C}_{\rm e}$. On a fixed time, this configuration is the same as the one given in Eq. \eqref{def-cv}, with $\chi$ being a multivalued angle with respect to the loop $\gamma$. The $N$-matching configuration corresponds to $a_{\beta_1}(\gamma_1) + \dots + a_{\beta_N}(\gamma_N)$\;, where $\gamma_1, \dots, \gamma_N$ are open vortex lines with common endpoints. Finally, as discussed in Ref. \cite{wavefunctional},  the gauge field of the 
nonoriented thin configuration is related to
\begin{align}
a(\{\gamma\}) = a_{\beta}(\gamma) + a_{\beta'}(\gamma') + a_{\mathscr{E}}(\delta)\;,\quad {\mathscr{E}} = \beta - \beta' \;,
\label{epsilon}
\end{align}
by means of a gauge transformation $S_{\rm D}$ which is singular but single-valued, so that it is irrelevant for Wilson loop computations. Therefore, the Abelian-projected version of the nonoriented vortex is given by vortex lines $\gamma$ and $\gamma'$ with magnetic charge $\beta$ and $\beta'$, respectively. Conservation of the flux is ensured due to a Dirac string along $\delta$. To get rid of the unobservable string $\delta$, while keeping the flux collimated, the wavefunctional was expressed in terms of a monopole potential $\zeta$:
\begin{align}
 \Psi (A,\zeta) = \sum_{\{\gamma\}} \psi_{\{\gamma\}}\, \delta\big(A - a(\{\gamma\})\big)\, \delta\big(\zeta - (-\Delta)^{-1}\nabla \cdot b(\{\gamma\})\big) \;,
 \label{b4}
\end{align}
where $b(\{\gamma\})$ is a superposition 
of thin fluxes $b_\beta(\gamma)=2\pi\beta \cdot T j(\gamma)$ and $\psi_{\{\gamma\}}$ is a phenomenological weight that incorporates center-vortex tension and stiffness. Changing to the (dual) electric-field representation, by means of a functional Fourier transform, we showed that the sum over vortex networks in Eq. \eqref{b4} lead to an effective field theory. This theory contains $N$ complex scalar fields $\phi_1, \dots, \phi_N$, corresponding to the charges $\beta_1, \dots, \beta_N$. The mixed percolating ensemble gives place to discrete $Z(N)$ vacua, which support a domain wall sitting on the Wilson loop. In this manner, an area law with asymptotic Casimir scaling was obtained.
\begin{figure}[h]
  \hspace{1.7cm}
  \includegraphics[ scale=.16]{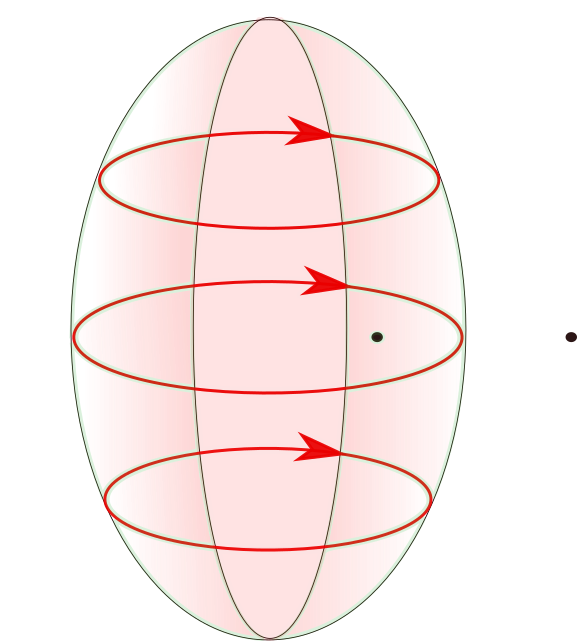}  
  \hspace{2.4cm}  
   \includegraphics[scale=.16]{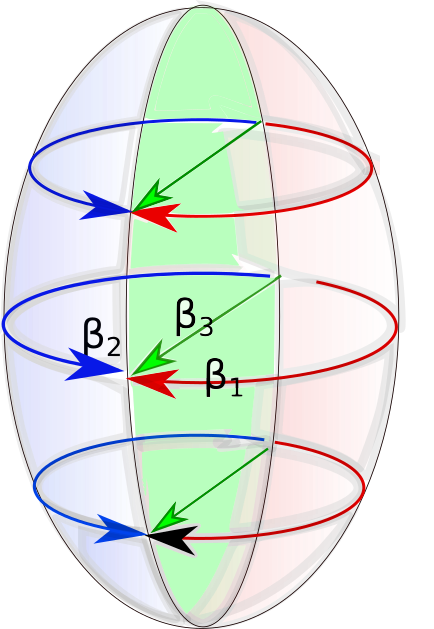}
   \hspace{2.4cm}  
     \includegraphics[scale=.16]{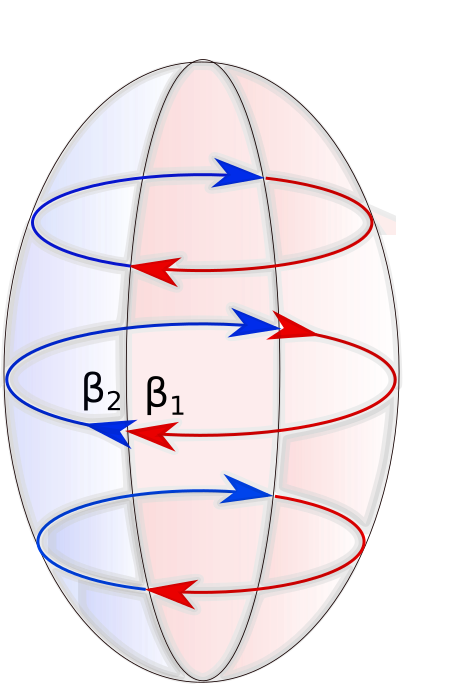}
     \caption{From left to right: worldsurface linking a Wilson loop (black dots); worldsurfaces spanned by $N$-matching in $SU(3)$, and by the nonoriented chain.}\label{vortices-4d}
\end{figure} 
The configurations in the wave functional represent the center-vortex surfaces as seen at a fixed time. Then,  in Euclidean spacetime, we considered the gauge field configurations
\begin{equation}
    a(\{\mathcal{S}\}) = \sum_{\beta, \mathcal{S}} a_\beta(\mathcal{S})\;,
\end{equation}
where $\{\mathcal{S}\}$ is a center-vortex network in $4d$ formed by vortex surfaces carrying a distribution of defining magnetic weights $\beta_i$. Explicit expressions for $a_\beta(\mathcal{S})$ were first given in Ref. \cite{cv-3}. The $N$-matching configuration can be described by $a_{\beta_1}(\mathcal{S}_1)+\dots+a_{\beta_N}(\mathcal{S}_N)$ \cite{weingarten-new}, where $\mathcal{S}_i$ are open worldsurfaces, with
    $\beta_1 + \dots + \beta_N = 0$ and  $\partial \mathcal{S}_1 = \dots = \partial \mathcal{S}_N$.
Regarding the collimated chains, in addition to the observable guiding centers of physical center vortices denoted by $\{\mathcal{S}\}$, the Abelian-projected fields also contain unobservable Dirac worldsheets $\mathcal{U}$. In particular, the $4d$ version of the nonoriented chain given in Eq. \eqref{epsilon} is  
\begin{gather}
a_{\beta}(\mathcal{S}) + a_{\beta'}(\mathcal{S}')+a_{\mathscr{E}}(\mathcal{U})\makebox[.2in]{,} 
    \mathscr{E} = \beta - \beta' \makebox[.2in]{,}   -\partial \mathcal{S} = \partial \mathcal{S}' =\partial\mathcal{U}   \;,
    \label{nontrivial_boundary}
\end{gather}
where $\mathcal{S}, \mathcal{S'}$, $\mathcal{U}$ are the surfaces spanned in time by $\gamma, \gamma', \delta$, respectively. In this manner, only configurations with collimated flux are considered (see Fig. \ref{vortices-4d}). For a general vortex network, the Wilson loop is given by 
\begin{equation}
    W_{\mathcal{C}_{\rm e}} = e^{i\Phi}\makebox[.5in]{,}\Phi = \sum_{\mathcal{S} \in \{\mathcal{S}\}} 2\pi\beta \cdot T  \, I(\mathcal{S},\mathcal{S}_{\rm e}),
\end{equation}
where $I(\mathcal{S},\mathcal{S}_{\rm e})$ is the intersection number between $\mathcal{S}$ and $\mathcal{S}_{\rm e}$. That is, the Dirac worldsheets do not contribute, as expected.  In the ensemble averages, we incorporated a phenomenological weight $\psi_{\{\mathcal{S}\}}$ to capture vortex properties observed in the lattice. In this manner, the Wilson loop average was modeled by
\begin{align}
    \langle W_{\mathcal{C}_{\rm e}}\rangle = \sum_{\{S\}} \psi_{\{S\}}\, e^{i\Phi}.
\end{align}
The phenomenological weights $\psi_{\{S\}}$ parametrize the simplest properties of center-vortex worldsurfaces and monopole worldlines by means of their respective tension parameters. They also measure the importance of the different interactions and connections among the components. 

\section{Closed surfaces and loops \`a la Weingarten}
\label{alaW}

As shown in Ref. \cite{weingarten}, a sum over noninteracting closed surfaces, with tension and a particular cost that depends on the Euler characteristics, can be regularized in the lattice by means of a Gaussian matrix model. The associated partition function can be interpreted as due to surfaces  $\mathcal{S}_{\rm c}$ with $N$ possible labels at each vertex (see Ref. \cite{weingarten-new}). More precisely ($F$ is the number of faces),
\begin{equation}
    Z_0 = \sum_{ \mathcal{S}_{\rm c}} e^{- \mu_0  A(\mathcal{S}_{\rm c})}\makebox[.5in]{,}A(\mathcal{S}) = a^2 F,
    \label{gene}
\end{equation}
is equivalent to a model with variables \( V(x,y) \in \mathbb{C}^{N \times N} \) defined at the links:
\begin{gather}
    Z_0=\int DV\, \exp\left(\gamma\sum_{p}{\rm Tr}V(p)-Q_0[V]\right)\makebox[.5in]{,}\mathcal{Q}_0[V] =   \sum_{\{ x,y\}} \mathcal{Q}_0(V(x,y))
    \nonumber \\
    \mathcal{Q}_0(V) = {\rm Tr} \left(  V^\dagger V \right) \makebox[.5in]{,}\ln\gamma^{-1}=\mu_0 a^2\;,
    \label{wein}
\end{gather}
Here, $V(p)$ is a plaquette variable given by the ordered product of the links that form the plaquette $p$. In the Abelian-projected setting, the center-vortex components are characterized by global weights. Those carrying a given magnetic weight $\beta$ will be generated by a complex field $V(x,y)$ ($N=1$). The general $N$ case is important to revisit the ensemble with non-Abelian d.o.f. (see Sec. \ref{nae}). As noted in Ref. \cite{weingarten}, the model defined by Eq. \eqref{wein} is pathological due to a divergent $Z_0$, even on a finite lattice with periodic boundary conditions. In that  reference, the model was stabilized with sixth or higher order interactions. In Ref. \cite{weingarten-new}, we considered a quartic model, replacing $\mathcal{Q}_0 \to \mathcal{Q}(V) = {\rm Tr}\left( \eta V^\dagger V+\lambda (V^\dagger V)^2\right) $. This generates interacting closed surfaces with a renormalized tension. Because of the Von Neumann trace inequality,  this turns out to be stable for $\lambda > 3 \gamma$. The action for the interacting partition function $Z$ can be written in the form
\begin{align}
& Q[V] - \gamma\sum_{p}{\rm Tr}V(p)=  3 \gamma\sum_{\{x,y\}}  {\rm Tr} ( V^\dagger (x,y)V(x,y))^2  -\gamma\sum_{p}{\rm Tr}V(p) \label{smem} \\
& +\lambda'  \sum_{\{x,y\}} {\rm Tr} \,  (V^\dagger(x,y) V(x,y) -\vartheta^2 I)^2    + c
\makebox[.5in]{,} \vartheta^2 = \frac{-\eta}{2\lambda'} \makebox[.3in]{,}
\lambda' = \lambda -3\gamma \label{sline} \;.
\end{align}
  By construction, the second member in line \eqref{smem} is positive definite. This also applies to the squared terms in line \eqref{sline}. When the renormalized tension is below a critical value  $\mu_{\rm c} >0$ (percolating surfaces), the parameter $\eta $ becomes negative, thus leading to a transition where $\vartheta^2$ becomes positive.  In the thermodynamic limit, a condensate dominated by the ``Goldstone'' modes
\begin{align}
    V(x,y) = \vartheta\,  U(x,y) \makebox[.5in]{,} U(x,y) \in U(N)  \;.
    \label{freeze}
\end{align}
is expected to be formed, accompanied by fluctuations with ``mass'' $\lambda' \vartheta^2$, which are suppressed if
$\lambda'$ is much larger than $\gamma$. In this regime, we can effectively freeze the ``modulus'' of $V(x,y)$ and retain only the softer Goldstone modes, which are governed by $\approx \gamma \vartheta^2   \sum_{p}{\rm Tr} (I - U(p) ) $, corresponding to the Wilson action for the $U(N)$ gauge field theory.
  This is the higher dimensional version of similar well known properties in ensembles of loops. In the lattice, consider the partition function for
  a gas of holonomies $\Gamma_R(\mathcal{C}) = R(x_1,x_2)\dots  R(x_{n},x_1)  $ defined over noninteracting loops $\mathcal{C}$ with $V$ vertices,
\begin{equation}
    \tilde{Z}_0 = \sum_{ \mathcal{C}}  e^{-\tilde{\mu} (L(\mathcal{C}_1)+ L(\mathcal{C}_2) + \dots)} \, {\rm Tr}\, \Gamma_R(\mathcal{C}_1)\,{\rm Tr}\, \Gamma_R(\mathcal{C}_2) \dots \makebox[.5in]{,}L(\mathcal{C}) = a V,
    \label{gene}
\end{equation}
where $R(x,y)$ is a $\mathcal{D} \times \mathcal{D}$ matrix source ($R(y,x) = R(x,y)^\dagger$). $\tilde{Z}_0$ can be rewritten by introducing a tuple $\zeta$ formed by $\mathcal{D}$ complex scalars:
\begin{gather}
    \tilde{Z}_0 = \int D\zeta\, \exp\left( \sum_{l} \tilde{\gamma} \,\tilde{V}(l)  -\tilde{Q}_0[\zeta]\right)\makebox[.5in]{,}  \tilde{Q}_0[\zeta] =   \sum_x \tilde{\mathcal{Q}_0}(\zeta(x))
 \;, \nonumber \\
\tilde{\mathcal{Q}}_0(\zeta)  =  \zeta^\dagger \zeta  \makebox[.5in]{,}   \ln \tilde{\gamma}^{-1} = \tilde{\mu}_0 a\;, 
\end{gather}
where $\tilde{V}(l) = \zeta^\dagger(x) R(x,y)\zeta(y)$ and the sum is over the oriented links $l=(x,y)$. As usual, a repulsive interaction among loops, $\tilde{\mathcal{Q}}_0 \to \tilde{\mathcal{Q}}(\zeta)  =   \tilde{\eta} \,  \zeta^\dagger \zeta +
\tilde{\lambda}\,  (\zeta^\dagger \zeta)^2$, renormalizes the tension ($\tilde{\mu}_0 \to \tilde{\mu}$) and allows for the stabilization of the boson condensate. When, $R^\dagger(x,y) R(x,y) = I_{\mathcal D} $, the interacting action in the partition function becomes
\begin{gather}
       \tilde{Q}[\zeta] -  \sum_{l} \tilde{V}(l)  =    \tilde{\gamma}\sum_{x,\mu}  (\Delta_\mu \zeta)^\dagger \Delta_\mu \zeta +  \sum_x
    \left( a^2 m^2  \zeta^\dagger \zeta 
    + \tilde{\lambda}\,  (\zeta^\dagger \zeta )^2 \right)
    \;, \nonumber \\
    \Delta_\mu \zeta =  R(x, x+ \mu) \zeta(x+ \mu) - \zeta(x) \makebox[.5in]{,}  \tilde{\eta} =   2d\tilde{\gamma} + a^2 m^2 \;,
    \label{Rg}
\end{gather} 
where $d$ is the spacetime dimension. The parameter $m^2$ determines whether spontaneous symmetry breaking occurs, with boson condensation at $\tilde{\mu} < \tilde{\mu}_{\rm c}$. Finally, rescaling $\zeta(x) \to a\zeta(x)$, the action in the continuum is obtained when $a\to 0$.

\section{Abelian-projected  oriented and nonoriented center vortices}

Taking into account the $N$ possible magnetic charges for the vortex worldsurfaces, the necessity of a quartic term to stabilize the model, and the $N$-matching configurations, the Wilson loop average over oriented center vortices, for quarks in representation ${\rm D}(\cdot)$, is given by
\begin{gather}
  \langle W_{\mathcal{C}_{\rm e}}\rangle_{\rm c.v.}=   Z_{\rm c.v.}[b] \propto\int DV \exp \left( -W_{\rm c.v.}[V] \right) \makebox[.5in]{,} b(p) = (2\pi k/N) s(p)\;, \label{bp}\\
    W_{\rm c.v.}[V] = -\gamma\sum_{p}{\rm Tr} \big( e^{i b(p)} V(p) \big) + Q[V] - \sum\limits_{\{x,y\}} \xi \big(\det V(x,y) + \det V(y,x)\big).\label{abelianvortices}
\end{gather}
Here, \(V(x,y)\) is an \(N \times N\) diagonal matrix whose entries are the $V_i(x,y) \in U(1)$ link variables, which represent vortex branches with global magnetic charges $\beta_i$. The $\xi$ term introduces configurations with $N$-matching, as it allows for $N$ open surfaces carrying weights $\beta_1,\dots,\beta_N$ to join at common links. Moreover, $b(p)$ is nontrivial when the $p$ plaquette intersects 
$\mathcal{S}_{\rm e}$ and $k$ is the $N$-ality of ${\rm D}(\cdot)$.  In such cases, $s(p)$ equals $+1$ ($-1$) depending on whether the relative orientation between $p$ and 
$ S_{\rm e} $ at the intersection point is positive (negative). Every time $\mathcal{C}_{\rm e}$ is linked, the frustration $e^{i b(p)}$ generates a center element. These are the fingerprints of center vortices.  

\subsection{Abelian-projected chains}
\label{abe-chains}

The mixed ensemble of center vortices and chains is given by 
\begin{gather}
\langle W_{\mathcal{C}_{\rm e}}\rangle_{\rm mix} = Z_{\rm mix}[b]\propto \int    DV D\phi\, \exp 
 \left( -W_{\rm mix}[V, \phi]\right) \nonumber \\
W_{\rm mix}[V, \phi] =W_{\rm c.v.}[V] - \sum_{l} \tilde{V}(l) +\tilde{Q}[\phi] \;, \nonumber \\
\tilde{V}(l) = \sum_{\alpha}\bar{\phi}_{\alpha}(x) H_\alpha(x,y) \phi_{\alpha}(y) \makebox[.5in]{,}H_\alpha(x,y) = \bar{V}_j(x,y)V_i(x,y)  \;,\nonumber \\
\tilde{Q}[\phi] =   \sum_x \sum_{\alpha} \tilde{\mathcal{Q}}(\phi_{\alpha}(x)) 
\makebox[.5in]{,} \tilde{\mathcal{Q}}(\phi)  =   \tilde{\eta} \,  \bar{\phi} \phi +
\tilde{\lambda}\,  (\bar{\phi} \phi)^2\;.
\label{abz}
\end{gather} 
 In effect, the expansion in powers of $\tilde{V}$ generates products of loop holonomies for the link variables $H_\alpha$, which, in a subsequent $\gamma$ expansion, allows for plaquettes of type $i$ to be glued with plaquettes of type $j$ along the loop. This is precisely a chain configuration, where an open vortex surface carrying charge $\beta_i$ is glued with another open surface carrying $\beta_j$ along a monopole loop carrying the magnetic charge $\beta_i-\beta_j$. 

\subsection{Percolating phase}
\label{abelian-percolating}

As discussed in Ref. \cite{weingarten-new}, when 
center vortices percolate and there is a strong effect of $N$-matching, the link variables have the form $V(x,y)=\vartheta U(x,y)$, with $U(x,y)=e^{i\theta(x,y)\cdot T} \in U(1)^{N-1}$, thus leading to
\begin{gather}
W_{\rm mix}[V, \phi] \approx  \gamma \vartheta^2   \sum_{p}{\rm Tr} \Big( I -  e^{i b(p)}  U(p) \Big) +  \sum_{x,\mu, \alpha} \vartheta^2(\Delta_\mu \phi_\alpha)^\dagger \Delta_\mu \phi_\alpha + \mathcal{U}(\phi) 
     \;, \nonumber \\
      \Delta_\mu \phi_\alpha =  R_\alpha(x, x+ \mu) \phi_\alpha(x+ \mu) - \phi_\alpha(x) \;, \quad \mathcal{U}(\phi)= \sum_{x,\alpha}  
    \big(a^2 m^2  |\phi_\alpha|^2 
    + \tilde{\lambda}\,  |\phi_\alpha|^4 \big)   \;,
     \label{Wmix}
\end{gather} 
  with $R_{\alpha}(x,y)=U_i(x,y)\bar{U}_j(x,y)$.
  
\subsection{Abelian Projection: $N$-ality and the Confining Flux Tube}
\label{Abe-ft}

To investigate the properties of the confining flux according to the Abelian-projected mixed ensemble, we considered a Wilson loop along a rectangular contour $\mathcal{C}_{\rm e}$ in the $x_0, x_1$ plane, formed by the lines $(\tau,\pm L/2,0,0),\, \tau \in (-\infty,\infty)$ along with two other lines at $x_0=\pm \infty$, which we neglected. The frustration in this setup is nontrivial on plaquettes that intersect $\mathcal{S}_{\rm e}$, which we chose to be $|x_1| > L/2, \, \forall x_0$, with $x_2=x_3 =0$. At an intermediate time slice ($x_0 =0$), there is a three dimensional cubic lattice. The quark and antiquark are localized at the centers of cubes positioned at $x_1 = \pm L/2$, $x_2=0$, $x_3 =0$. The partition function $Z_{\rm mix}[b]$ is dominated by the lowest action configurations of $W_{\rm mix}[V, \phi]$, which could be studied through lattice simulations. Instead, we studied these solutions in the continuum limit. The central issue is determining which configuration minimizes the action. Naturally, the smallest action corresponds to the case where the flux tube guiding center is  a straight line. The optimal magnetic weight $\beta'$ can be found by minimizing the action
\begin{gather}
    S(J')=\int d^4x \left[ \frac{\gamma\vartheta^2}{4}(F_{\mu\nu}-J'_{\mu \nu})^2 +\sum_\alpha  D_\mu \bar{\phi}_\alpha D_\mu  \phi_\alpha+ \mathcal{U}(\phi)\right]\makebox[.5in]{,}J'_{\mu\nu}=2\pi\beta'\cdot T s_{\mu\nu}(\mathcal{S}_{\rm e})\;,
\end{gather}
where all possible $\beta'$ for $SU(N)$ representations with $N$-ality $k$ have to be considered, as they all generate the same center element in the lattice description.\footnote{This is in sharp contrast with the monopole-only scenario, which also gives rise to this action, but with a minimization done by only considering the magnetic weight of the quark representation.} In general, the $k$-Antisymmetric representation is expected to give the lowest action. Indeed, this was shown to be the case at the BPS point \cite{weingarten-new}. Then, we also showed that this model displays an asymptotic Casimir law, i.e. $\sigma_k\propto k(N-k)\, \sigma_1$, where $\sigma_k$ is the asymptotic tension for a quark representation with $N$-ality $k$.

\subsection{Double-winding Wilson loops}

\begin{figure}[h]
    \centering
    \begin{subfigure}[b]{0.4\textwidth}
\includegraphics[width=\textwidth]{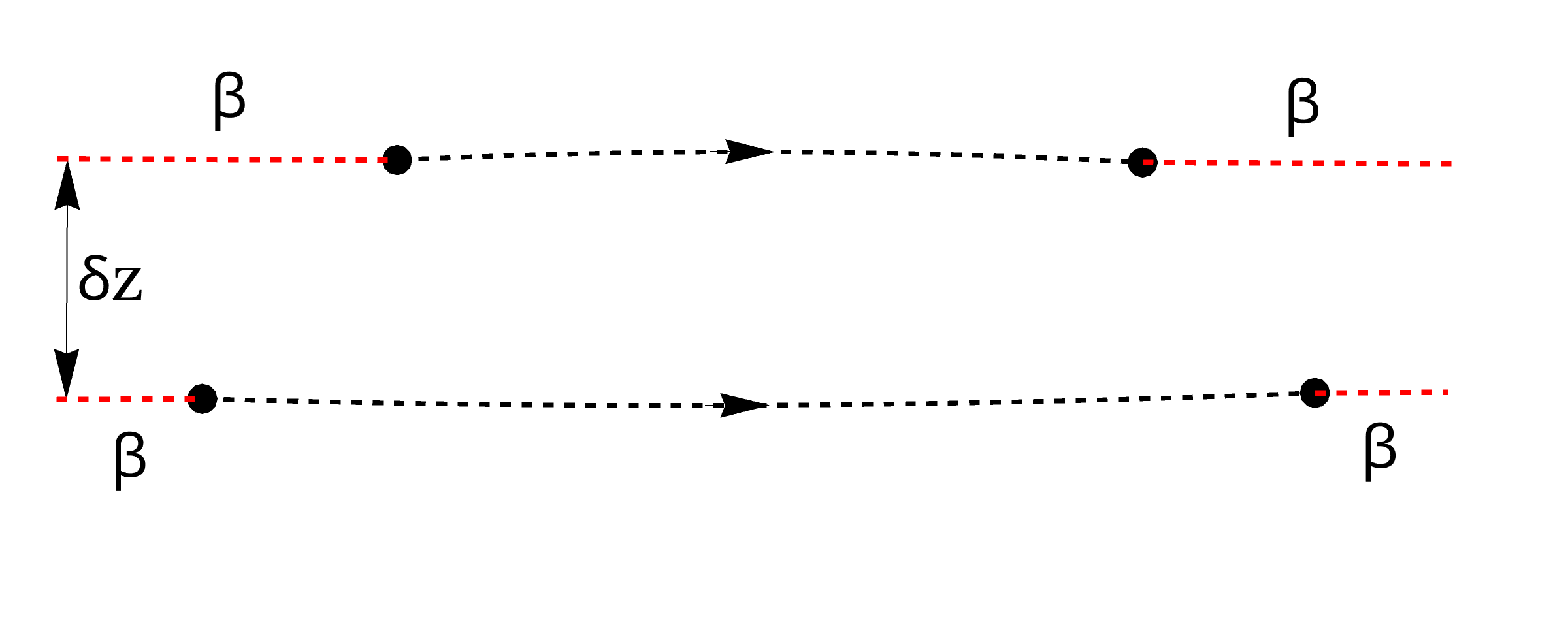}
    \caption{Monopole-only scenario.}
     \label{fig:mon-only}
    \end{subfigure}
\hspace{1.7cm}    \begin{subfigure}[b]{0.4\textwidth}
\centering
\includegraphics[width=\textwidth]{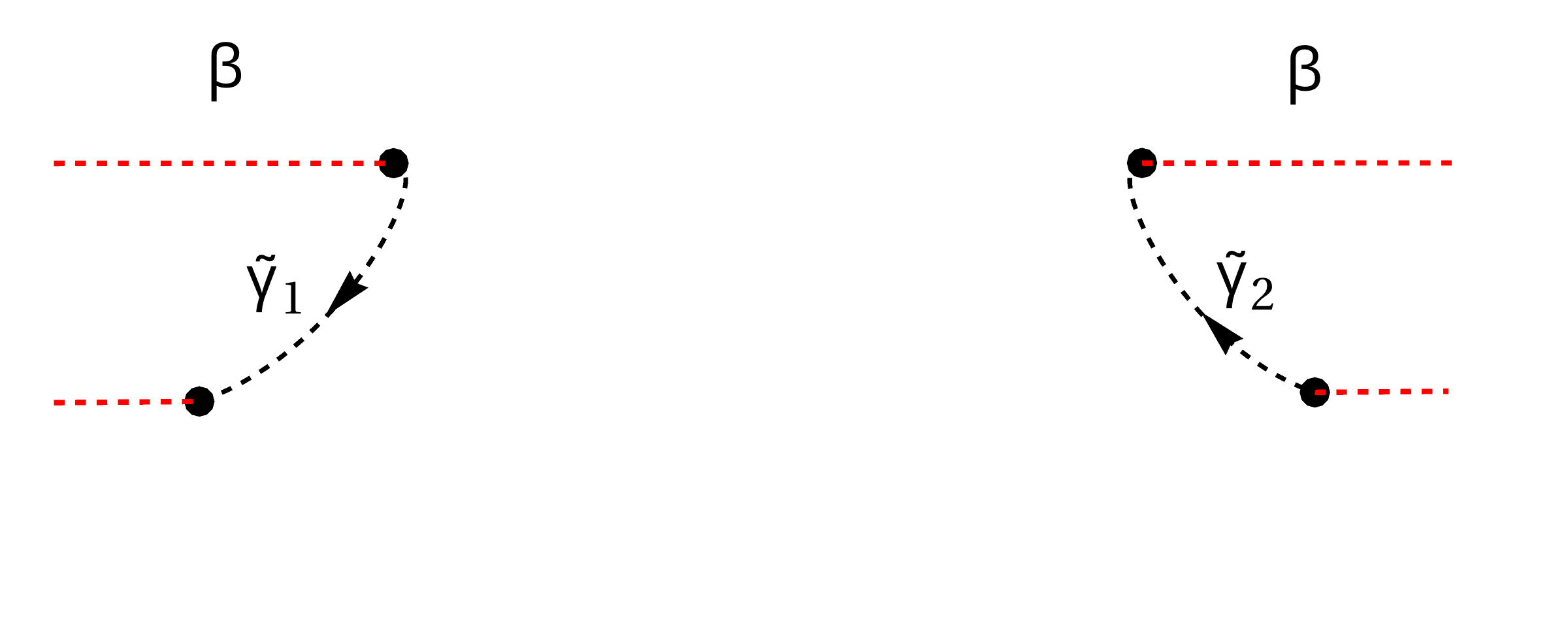}
    \caption{Collimated chains: $SU(2)$.}
    \label{fig:chains-su2}
    \end{subfigure}
    \begin{subfigure}[b]{0.4\textwidth}
        \centering
\includegraphics[width=\textwidth]{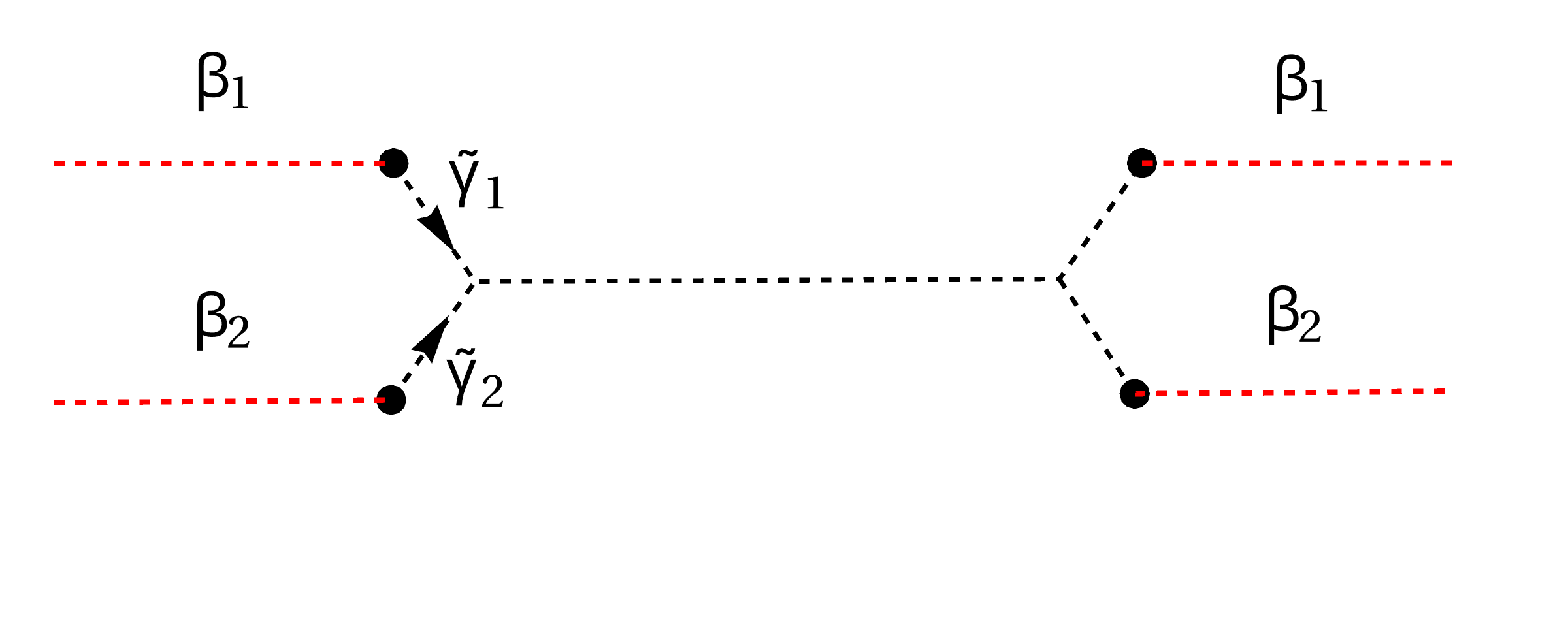}
    \caption{Collimated chains: $SU(3)$.}
    \label{fig:chains-su3}
\end{subfigure}
\caption{Double-winding Wilson loops: the dotted red lines are the surface $\mathcal{S}_{\rm e}$ at a fixed time.}
\end{figure}

In Ref. \cite{weingarten-new}, the double-winding Wilson loop for quarks in the defining representation of $SU(N)$, was discussed for the Abelian-projected mixed ensemble of center vortices and chains, and also for the monopole-only ensemble. This observable was introduced in Ref. \cite{d-wl} as a way to clearly differentiate between the center-vortex and the monopole-only scenarios. In $SU(N)$, the monopole-only effective action is given by $S(J)$, where $J_{\mu\nu} $ carries a magnetic weight $\beta$ of the defining representation. This type of monopole-only scenario was studied in Ref. \cite{maedan} and reviewed in \cite{yanddelta}. At a given intermediate time, the magnetic charge $\beta$ is carried all along the double Wilson loop, leading to the sources represented in Fig. \ref{fig:mon-only}. In this case, the lowest action will be attained by a pair of flux tubes carrying flux $\beta$, which could interact as $\delta z$ becomes small. Therefore, the Wilson loop will be approximately given by the sum of the areas spanned by these flux tubes. On the other hand, for the mixed ensemble of collimated objects, all possible distributions of weights have to be considered.  When $\delta z$ is small,  the lowest action in the continuum limit will be characterized by the angles $\xi_1$, $\xi_2$, which change by $2\pi$ when going around the paths $\tilde{\gamma}_1$, $\tilde{\gamma}_2$,  shown for $SU(2)$ and $SU(3)$ in Figs. \ref{fig:chains-su2} and \ref{fig:chains-su3}, respectively. For $SU(2)$, the flux tubes will be on top of the curves $\tilde{\gamma}_1,\tilde{\gamma}_2$ depicted in Fig. \ref{fig:chains-su2}, which implies a difference-of-areas law. For $SU(3)$, as $\beta_1+\beta_2=-\beta_3$, the flux tubes merge into a single one, leading to the double-Y shaped configuration in Fig. \ref{fig:chains-su3}. Moreover, this property implies that the flux tube between three quark sources is Y shaped, which is consistent with lattice simulations in Abelian-projected scenarios \cite{ab-proj-4}.

\subsection{The effect of monopole matching}

\begin{figure}[h]
    \centering
    \begin{minipage}{0.45\textwidth}
        \centering
        \includegraphics[scale=0.16]{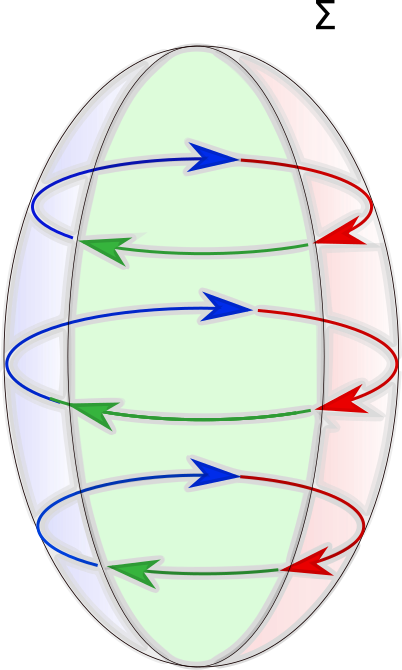}
    \end{minipage}
    \hspace{1cm}
    \begin{minipage}{0.45\textwidth}
        \centering
        \includegraphics[scale=0.16]{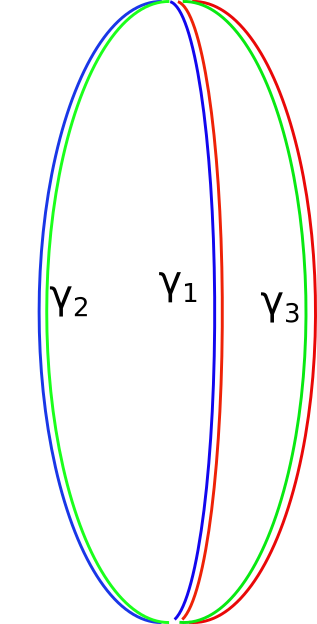}
    \end{minipage}
       \caption{From left to right: worldsurfaces spanned by a nonoriented chain in $SU(3)$; matching between the corresponding monopole worldlines.}
    \label{fig:main}
\end{figure}

The matching of three monopole worldlines at a single point (see Fig. \ref{fig:main}) can be incorporated by introducing a term cubic in the scalar fields \cite{weingarten-new}. New quartic monopole interactions involving terms such as $|\phi_\alpha|^2 |\phi_\beta|^2$ can also be included. As a result, even within the Abelian-projected framework, the field profiles do not necessarily conform to the Nielsen-Olesen type, and deviations from the asymptotic Casimir scaling may occur. 

\section{ Revisiting the non-Abelian partition function}
\label{nae}

Another objective of Ref. \cite{weingarten-new} was to formulate  the proposal of Ref. \cite{mixed} in terms of the Weingarten matrix representation. In Ref. \cite{mixed}, the oriented center-vortex sector of the non-Abelian ensemble was considered as an extension of that for Abelian worldsurfaces discussed in Ref.  \cite{Rey}. In the Abelian case, a loop condensate is described by a Wilson action for a $U(1)$ gauge field that represents the Goldstone modes. Moreover, when these worldsurfaces are minimally coupled to a Kalb-Ramond field $b_{\mu\nu}$, a frustration $e^{ia^2b_{\mu\nu}}$ emerges. Based on these considerations, the Wilson loop average over oriented center vortices was directly proposed to be given by the Wilson action for non-Abelian link variables $U_\mu(x)$ with a frustration $e^{i b(p)} $, with $b(p)$ given in Eq. \eqref{bp}. The use of link variables in $SU(N)$ simply implemented the $N$-matching property among center vortices. Nonoriented configurations were included by means of adjoint holonomies. On this direction, our contribution in Ref. \cite{weingarten-new} was presenting a clear relationship between percolating center-vortex worldsurfaces with non-Abelian d.o.f. and the corresponding  gauge-field Goldstone modes.  In this case, the 4D center-vortex sector was generated by Eq. \eqref{abelianvortices}, but with a general complex matrix $V(x,y) \in \mathbb{C}^{N \times N}$.  Indeed, this way, the loops propagated along worldsurfaces are associated with non-Abelian Wilson loops \cite{weingarten}. As in the Abelian-projected case, after stabilizing the ensemble with contact terms, when center vortices percolate and the effect of $N$-matching is strong, $V(x,y)$ is forced to be close to $\vartheta U(x,y)$, where $U(x,y) \in SU(N)$. In addition, in Ref. \cite{weingarten-new}, the nonoriented component was generated by  adjoint scalars $\zeta_\alpha$ labeled by the positive roots of $SU(N)$.  Similarly to the models in Ref. \cite{mixed}, the system can undergo $SU(N) \to Z(N)$ SSB, but with a simplified content based on the simplest monopole matching rules \cite{prospecting-paper}.

\section{Discussion}
\label{conclusions}

In this work, we briefly reviewed the main results presented in Ref. \cite{weingarten-new}.  Abelian-projected variables derived from Monte Carlo configurations are known to capture $N$-ality at asymptotic distances \cite{ab-proj-1, ab-proj-2}, reproduce the flux tube for fundamental quarks \cite{ab-proj-3, suga2014}, and describe the three-quark potential in $SU(3)$ \cite{ab-proj-4}. Our findings successfully align with these observations, showing how an Abelian mixed ensemble of oriented and nonoriented center vortices can account for $N$-ality while supporting a ``dual superconductor'' model of confinement. 
 The main ideas parallel those previously proposed in Ref. \cite{mixed}, where the mixed ensemble components included non-Abelian degrees of freedom. When center vortices percolate and the monopoles on the nonoriented chains proliferate, a flux tube between quark probes was shown to be formed. In this respect, the Weingarten formulation of Ref. \cite{weingarten-new} clarified the emergence of the non-Abelian Goldstone modes when center vortices percolate. Both the Abelian-projected and non-Abelian settings can be formulated in terms of matrix models with diagonal and general link variables $V(x,y) \in \mathbb{C}^{N \times N}$, respectively. The diagonal variables generate center-vortex branches carrying global magnetic weights, while the general variables generate branches with local weights, which vary at each vertex.  Both settings exhibit  $N$-ality, Casimir law, Abelian transverse profiles which are $k$-independent, and provide results for double-winding Wilson loops which are in accordance with lattice simulations. They could also accommodate deviations from the exact Casimir law observed in recent lattice simulations \cite{latt-scaling}.  The $U(1)^{N-1}$ symmetric Weingarten matrix model for the Abelian-projected ensemble naturally embeds into the Cartan sector of the $SU(N)$ description. Then, depending on the phenomenological parameters, the non-Abelian model may behave as if it has global weights on the center-vortex branches. An interesting aspect of the non-Abelian scenario is its potential to incorporate additional features of confinement. Among them is the possibility of describing the correct physical picture for adjoint string breaking and establishing a smoother connection to confinement at intermediate distances, where Abelian projection proves inadequate \cite{noabelianadjointtube}. Indeed, center-vortex thickness and non-Abelian degrees of freedom and have been proposed to explain the intermediate Casimir scaling \cite{casimirscalingthickness,casimirscalingthickness-2}. In addition, since both settings interpolate between percolating and normal phases, it would be worthwhile to analyze whether they can capture the geometry of center-vortex worldsurfaces observed at finite temperatures \cite{leinweber-nmatching}.

\acknowledgments 
This work was partially supported by the Conselho Nacional de Desenvolvimento Cient\'ifico e Tecnol\'ogico (CNPq), under contract 309971/2021-7, and the S\~ao Paulo Research Foundation (FAPESP), grant no.
2023/18483-0.

\end{document}